# Numerical study of the Steady State Fluctuation Relations Far from Equilibrium.

Stephen R. Williams[a], Debra J. Searles[b] and Denis J. Evans[a]

Research School of Chemistry, Australian National University, Canberra, ACT 0200, AUSTRALIA

Nanoscale Science and Technology Centre, School of Science, Griffith University, Brisbane, Qld 4111, AUSTRALIA

16/01/06

## Abstract

A thermostatted dynamical model with five degrees of freedom is used to test both the Evans-Searles and the Gallavotti-Cohen fluctuation relations (ESFR and GCFR respectively). In the absence of an external driving field, the model generates a time independent ergodic equilibrium state with two conjugate pairs of Lyapunov exponents. Each conjugate pair sums to zero. The fluctuation relations are tested numerically both near and far from equilibrium. As expected from previous work, near equilibrium the ESFR is verified by the simulation data while the GCFR is not confirmed by the data. Far from equilibrium where a positive exponent in one of these conjugate pairs becomes negative, we test a conjecture regarding the GCFR made by Gallavotti and co-workers. They conjectured that where the number of nontrivial Lyapunov exponents that are positive becomes less than the number of such negative exponents, then the form of the GCFR needs to be corrected. We show that there is no evidence for this conjecture in the empirical data. In fact as the field increases, the uncorrected form of the GCFR appears to become more accurate. The real reason for this observation is likely to be that as the field increases, the argument of the GCFR more and more accurately approximates the argument of the ESFR. Since the ESFR works for arbitrary field strengths, the uncorrected GCFR appears to become ever more accurate as the field increases. The final point of evidence against the conjecture is that when the smallest positive exponent changes sign, the conjecture predicts a discontinuous change in the "correction factor" for GCFR. We see no evidence for a discontinuity at this field strength; only a gradual improvement of degree of agreement as the field increases.

**Introduction**

Steady State Fluctuation Relations (SSFRs) describe the statistical fluctuations in time-averaged properties of nonequilibrium steady state dynamical systems. They show how thermodynamic irreversibility emerges from the time-reversible dynamics of the particles, and thus are of fundamental importance. The relationships also make quantitative predictions about these fluctuations, and these have been tested in computer simulations (for example see references [1-20]) and in laboratory experiments [21-24].

A number of different classes of Fluctuation Relations (FRs) have been proposed for deterministic, reversible dynamics. Transient Fluctuation Relations describe the statistics of time-averaged properties along a set of trajectory segments all initiated from a known distribution function at $t = 0$. For systems with the time-averaging commencing at $t = 0$, they can be written:

$$\frac{1}{t} \ln \frac{\Pr(\bar{\Omega}_t = A)}{\Pr(\bar{\Omega}_t = -A)} = A \tag{1}$$

where $\bar{\Omega}_t t = \int_0^t \Omega(s)ds \equiv ln[f(\Gamma(0),0)/f(\Gamma(t),0)] - \int_0^t ds\, \Lambda(s)$, and $\Omega$, the dissipation function, is a generalized entropy production that is uniquely defined for a specified dynamical system $\Gamma(t)$, and initial distribution of states $f(\Gamma,0)$, [25]. The notation $Pr(\bar{\Omega}_t = A)dA$ is used to represent the probability that $\bar{\Omega}_t$ takes on a value $(A - dA/2, A + dA/2)$. These relations have been derived for reversible deterministic systems that satisfy the ergodic consistency condition [25]. They are valid at all times, and do not explicitly require the dynamics to be chaotic. Transient Fluctuation Relations, are derived using the time reversal symmetry of the dynamics and hence apply to systems that are arbitrarily far from equilibrium. Of course in such systems the probability of negative values for the time averaged dissipation function become rather small, necessitating either short observation times or small system sizes. As written (1), these TFTs are formally ensemble $f(\Gamma,0)$, and dynamics $\Gamma(t)$, independent, although the precise expression for the dissipation function will change with different ensembles and dynamics.

Historically the first FRs that were proposed [26] concerned fluctuations in time averaged entropy production in nonequilibrium steady states, where trajectory segments were sampled from the single, unique steady state trajectory. The first SSFR was proposed for isoenergetic steady state systems and can be expressed

$$\lim_{t \to \infty} \frac{1}{t} \ln \frac{Pr(-\bar{\Lambda}_t = A)}{Pr(-\bar{\Lambda}_t = -A)} = A \qquad (2)$$

where t is the averaging time, $\bar{\Lambda}_t = \frac{1}{t}\int_0^t \Lambda(s)ds$ and $\Lambda$ is the phase space contraction rate $\Lambda = \frac{\partial}{\partial \mathbf{\Gamma}} \cdot \dot{\mathbf{\Gamma}}$ [27]. Following the early work of Evans, Cohen and Morriss (ECM2), [26] a formal derivation of this SSFR was given by Gallavotti and Cohen [28, 29] under the condition that A is bounded by a value $A^*$: $A \in (-A^*, A^*)$ [40]. ECM2 considered only isoenergetic dynamics but the work of Gallavotti and Cohen seemed to allow the application of the SSFR to a much wider class of dynamics (e.g. constant temperature dynamics). Evans and Searles [4, 6, 25] addressed the issue of SSFRs for steady states that are not maintained at constant energy. They gave a heuristic proof backed up by extensive numerical data that in steady states which are unique (i.e. steady state properties are independent of the initial phase) the dissipation function satisfies the SSFR

$$\lim_{t \to \infty} \frac{1}{t} \ln \frac{\Pr(\bar{\Omega}_t = A)}{\Pr(\bar{\Omega}_t = -A)} = A. \qquad (3)$$

This expression is derived from the corresponding TFT (1). The definition of the dissipation function depends on the initial ensemble and the details of the time reversible equations of motion but the form of equation (1) does not. In contrast to the Gallavotti-Cohen proof of (2) the proof of (3) requires no bounds on the values of $A$. Eq. (3) is expected to be valid for any suitable dynamics (constant energy, constant temperature, constant pressure). In the particular case of constant energy dynamics (3) is *identical* to (2) since $\Omega(t) = -\Lambda(t)$ but for other dynamics (such as thermostatted dynamics), $\Omega(t) \neq -\Lambda(t)$ instantaneously and therefore (2) and (3) are not equivalent.

For thermostatted field-driven non-equilibrium steady states that satisfy the condition of adiabatic incompressibility of phase space [30, 31],

$$\Omega(t) = -\beta J(t) V F_e \quad (4)$$

where $J$ is the dissipative flux, $V$ is the volume, $F_e$ is the (constant) applied field, $\beta = 1/(k_B T)$, $k_B$ is Boltzmann's constant and $T$ is the temperature of the thermostat. In this case, (3) is expected to apply for all observable values of $A$. Equation (4) shows that in the linear regime close to equilibrium the average dissipation function is indeed the spontaneous entropy production discussed in linear irreversible thermodynamics.

These FRs have been tested on various systems for example [1-20]. Equation (2) has been shown to apply in both the linear and nonlinear regime to isoenergetic systems, and equation (3) has been shown to apply in both the linear and nonlinear regime to a range of systems including isoenergetic, isokinetic and Nosé-Hoover thermostatted systems [1-10, 17, 19, 20, 32], and has recently be verified experimentally [24]. More recently Searles, Rondoni and Evans have presented a detailed mathematical proof of (3) for chaotic systems [31].

The Gallavotti and Cohen SSFR, namely (2), has only been validated numerically for constant energy dynamics. For constant temperature dynamics it has proved impossible to confirm (2) numerically, particularly for weak fields [6, 19, 20]. However, because (2) is an asymptotic relation it is always possible that the empirical data has not been considered at sufficiently long times for convergence to occur. The status of (2) for non-isoenergetic dynamics has recently been considered in detail by Evans, Searles and Rondoni [33].

To further complicate the issue, the formal derivation of (2) [28, 29], puts a limit on the magnitude of the external field. To derive (2) using the approach of Gallavotti and Cohen [28, 29] it is assumed that the dynamics is transitive, and in determination of (3) [25, 31] it is assumed that only a single steady state exists. Due to these requirements, it has been proposed that equation (2) might break down at large fields [11, 32], particularly when the transitive property is lost, as it is when there are unequal numbers of positive and negative exponents. In reference [11] Gallavotti and coworkers propose a modified version of (2), with a factor introduced to account for the reduction in dimensionality of the system as the dissipative field increases. This proposal results in a modification of (3),

$$\lim_{t\to\infty} \frac{1}{t} \ln \frac{\Pr(-\bar{\Lambda}_t = A)}{\Pr(-\bar{\Lambda}_t = -A)} = XA \qquad (5)$$

where X is equal to the ratio of the number of conjugate pairs of exponents where one exponent is positive and one is negative, divided by the number of conjugate pairs of nonzero exponents. We shall refer to (5) as GCX. The "correction factor" X is only expected to differ from unity at large fields. This in turn means that the X-factor cannot help the problems previously noted in confirming (2) for nonisoenergetic dynamics at weak fields.

In reference [32] Gallavotti et. al. propose the analogous modification to (3) namely

$$\lim_{t \to \infty} \frac{1}{t} \ln \frac{\Pr(\bar{\Omega}_t = A)}{\Pr(\bar{\Omega}_t = -A)} = XA \qquad (6)$$

We refer to equation (6) as GX, because although it refers to a modification of the ESFR, this modification was not proposed and is not supported by Evans and Searles. Indeed Evans and Searles have argued on many occasions that their relations (1,3) are exact as they are without any correction factors.

Here we carry out numerical tests to determine the value of X at large fields, and therefore determine whether or not (3) is valid (in which case X = 1). Although the arguments [11] that $X \neq 1$ are based on the behaviour of the phase space contraction, is seems reasonable to carry out this test since equation (2) and (3) become equivalent for isoenergetic systems. In this work we also determine the Lyapunov exponents for the system at each state point to identify if we are in a region where this factor would be expected to significantly alter the FR [11].

Testing (6) has been attempted in the past (e.g. [32]). However it is not easy to find systems where the Lyapunov exponents are "soft" so that they can change sign at fields that are sufficiently weak that negative fluctuations in the dissipation can still be observed. In the present paper we test (5) & (6) for thermostatted dynamics. For constant energy dynamics there is ample data showing that (2) and (3) are both valid at low to moderate field strengths when X = 1.

We test (3) on systems close to equilibrium, and also very far from equilibrium where the number of positive and negative exponents is not equal. We consider a dynamical system which is a variation on systems developed by Hoover and coworkers [34, 35] to model thermal conduction. For the equations and system parameters we choose, numerical results indicate it has a single steady state: the system is ergodic and strongly mixing so that the steady state is invariant to the initial configuration. We show that for this system, (3) can be verified, even far from equilibrium. We also test equation (2) for this system, although it is not expected to hold at small fields [19, 20, 33].

The dynamics for this system are not symplectic or µ-symplectic when the system is out of equilibrium [36], so we do not expect conjugate pairing of Lyapunov exponents, however we find that it is possible to drive the system so that the numbers of positive and negative exponents are unequal but negative fluctuations in the dissipation can still be observed. The system has five degrees of freedom, and therefore 5 Lyapunov exponents. One of these always has a value of zero since the dynamics is autonomous [37]. This system is suitable for directly testing the proposed modification of the FR [11] since the value of X proposed requires pairs of expanding and contracting directions close to equilibrium. In our case when close to equilibrium there will be 2 positive and 2 negative exponents corresponding to 2 pairs of expanding/contracting directions in phase space. When the system is driven sufficiently far from equilibrium, one of the positive exponents will become negative, reducing the number of expanding-contracting pairs from 2 to 1 with the factor $X$ being reduced from unity to $X = 1/2$. If this system is described by the chaotic hypothesis, and if the postulated modification to equation (3) is correct, then this would be clearly evident in a test of the FR.

**Model**

Hoover et al. [34] give a simple oscillator model for the nonequilibrium dynamics of heat flow. We have studied this model and accurately reproduced the results of Hoover et al. [34]; we have also found it to obey (3). The Hoover oscillator has only four degrees of freedom resulting in 3 nonzero Lyapunov exponents. The odd number of nonzero exponents makes it unsuitable for investigating (5) or (6) so we will not present these results here or discuss this model further. The model we use, given below, has a dissipation function that is different to the phase space compression factor. While the steady state *average* of the dissipation function (entropy production) and the steady state *average* of the phase space compression are equal, their distribution functions, which determine the fluctuations in these quantities, may well be different. We note that in general, it is not possible to derive (3) from (2) unless $\Lambda(t) = -\Omega(t)$ instantaneously, although this has been falsely assumed in the past. For this reason we have chosen dynamics where the dissipation function and phase space compression are different, to illustrate this elementary, yet unfortunately common error and hopefully help to demonstrate the importance of recognizing this difference and to clarify some of the confusion surrounding it.

The system we consider has three thermostatting terms and four non-trivial Lyapunov exponents allowing equations (5) & (6) to be tested. The equations of motion are:

$$\dot{q} = p$$
$$\dot{p} = -q - \alpha_1 p - \alpha_3 p^3 - \alpha_5 p^5$$
$$\dot{\alpha}_1 = \left(p^2 - T(q)\right)/\tau_1^2$$
$$\dot{\alpha}_3 = \left(p^4 - 3p^2 T(q)\right)/\tau_3^2 \quad (7)$$
$$\dot{\alpha}_5 = \left(p^6 - 5p^4 T(q)\right)/\tau_5^2$$

$$T(q) = 1 + \varepsilon \tanh(q)$$

where $q$ is the oscillator coordinate, $p$ is the momentum, $\alpha_1$, $\alpha_3$ and $\alpha_5$ are the multipliers which control the second, fourth and sixth moments of the momentum distribution, and $\tau_1$, $\tau_3$ and $\tau_5$ are the thermostat relaxation times. By setting $\varepsilon = 0$ we obtain the equilibrium equations of motion. Setting $0 < \varepsilon < 1$ results in a $q$ dependent temperature and the system is driven into a nonequillibrium steady state. The phase space compression factor, $\Lambda(p, \alpha_1, \alpha_3, \alpha_5)$ is given by,

$$\Lambda(p, \alpha_1, \alpha_3, \alpha_5) = \frac{\partial \dot{q}}{\partial q} + \frac{\partial \dot{p}}{\partial p} + \frac{\partial \dot{\alpha}_1}{\partial \alpha_1} + \frac{\partial \dot{\alpha}_3}{\partial \alpha_3} + \frac{\partial \dot{\alpha}_5}{\partial \alpha_5} = -\alpha_1 - 3\alpha_3 p^2 - 5\alpha_5 p^4 \quad (8)$$

Defining $H = H_0 + \frac{1}{2}\left(\tau_1^2 \alpha_1^2 + \tau_3^2 \alpha_3^2 + \tau_5^2 \alpha_5^2\right) = \frac{1}{2}\left(q^2 + p^2 + \tau_1^2 \alpha_1^2 + \tau_3^2 \alpha_3^2 + \tau_5^2 \alpha_5^2\right)$ (where $H_0$ is the Hamiltonian of the unthermostatted oscillator) we obtain $\dot{H} = -T(q)\Lambda(p, \alpha_1, \alpha_3, \alpha_5)$. At equilibrium we observe that $\dot{H} = \Lambda(p, \alpha_1, \alpha_3, \alpha_5)$ and use the Liouville equation to obtain the equilibrium distribution function for the system [30, 34],

$$f(q,p,\alpha_1,\alpha_3,\alpha_5) = \frac{\tau_1 \tau_3 \tau_5}{(2\pi)^{5/2}} \exp(-H(q,p,\alpha_1,\alpha_3,\alpha_5)). \tag{9}$$

We may now obtain the dissipation function from its definition [6, 25],

$$\bar{\Omega}_t t = \int_0^t ds\, \Omega(\mathbf{\Gamma}(s)) \equiv \ln\left[\frac{f(\mathbf{\Gamma}(0),0)}{f(\mathbf{\Gamma}(t),0)}\right] - \int_0^t \Lambda(\mathbf{\Gamma}(s))ds, \tag{10}$$

that is,

$$\Omega(q,p,\alpha_1,\alpha_3,\alpha_5) = (1-T(q))(\alpha_1 + 3p^2\alpha_3 + 5p^4\alpha_5) \tag{11}$$

This system was chosen because it is of low dimensionality, which means that the number of exponents is small and the relative imbalance in the number of positive and negative exponents is significant, even when there is only one additional negative exponent. The low dimensionality of the system also allows the phase space distribution to be visualized, and the precise determination of the Lyapunov exponents. Furthermore, the work of Hoover and coworkers show that their model (which is similar to ours) can be driven to a region where an imbalance in the number of exponents is obtained, and they have shown how the phase space distribution is altered [34].

**Simulations and Calculation of Lyapunov Spectra**

Following Hoover et al. [34] the equations of motion of the systems were solved using a fourth order Runge-Kutta algorithm. The time constants were set to $\tau_1 = 1, \tau_2 = 10, \tau_3 = 100$. A series of nonequilibrium systems were then studied with $\varepsilon = 0.1, 0.2, 0.3, 0.4, 0.43, 0.45$. Steady state simulations were performed, and the single long trajectory was divided into a large number of segments to form time-averages and then produce histograms of $\bar{\Omega}_t$ and $\bar{\Lambda}_t$ both close to equilibrium $\varepsilon = 0.1$ and far from equilibrium $\varepsilon = 0.43$. These distributions were then used to test Eq. (2) and (3).

The method used to calculate the Lyapunov spectra closely resembles that described in detailed by Dellago et al. [38] in their study of hard disk systems, and also used in reference [36]. To reduce numerical error, this method was modified to ensure that the zero exponent in the direction of the flow is identically zero as expected theoretically [37], i.e. no displacement of the tangent vectors in the direction of flow were allowed.

**Results and Discussion**

The Lyapunov spectra for various values of $\varepsilon$ are presented in Table 1. The exponent of the vector in the direction of flow in phase space is always zero and not included, leaving 4 nontrivial exponents. With $\varepsilon = 0$ we have two conjugate pairs (i.e. pairs that sum to zero) and an exponent that is identically zero. This is characteristic of an equilibrium state where the system has no preferred direction in time.

With $\varepsilon = 0.1$ the negative exponents are slightly larger in magnitude than their corresponding positive exponents and the system now evolves forward in time with an increasing probability of observing positive dissipation. As the time for which the trajectory segment is observed increases, the probability of observing positive dissipation increases as quantified by the fluctuation relation of equation (3), the ESFR. Under these weakly driven conditions it can be seen that the exponents conjugate pair around a nonzero value in the same way that $\mu$-symplectic dynamics would behave. For $\varepsilon = 0.43$ we have three negative exponents and a single positive one. The exponents under these strongly driven conditions no longer conjugate pair around a finite value, this is expected as the system is not $\mu$-symplectic. When the system is driven much harder it approaches stability where it will eventually follow a limit cycle in the steady state [34]. A stable system is characterized by the absence of positive Lyapunov exponents.

Table 1. Lyapunov spectra with the trivial exponent omitted.

| $\varepsilon$ | $\lambda_1$ | $\lambda_2$ | $\lambda_3$ | $\lambda_4$ | Error in exponents (~2 SE) |
|---|---|---|---|---|---|
| 0 | 0.0173 | 0.0025 | -0.0025 | -0.0173 | 0.0001 |
| 0.1 | 0.0195 | 0.0028 | -0.0032 | -0.0199 | 0.0001 |
| 0.2 | 0.0190 | 0.0018 | -0.0055 | -0.0226 | 0.0001 |
| 0.3 | 0.0131 | 0.0010 | -0.0089 | -0.0288 | 0.0001 |
| 0.4 | 0.0080 | 0.0008 | -0.0082 | -0.0320 | 0.0001 |
| 0.43 | 0.0063 | -0.0009 | -0.0088 | -0.0222 | 0.0001 |
| 0.45 | 0.00130 | -0.00400 | -0.01330 | -0.02310 | 0.00003 |

The fluctuation relations of equations (2) and (3) may be partially summed to obtain what is referred to as the integrated fluctuation relation (IFT) [2],

$$\lim_{t \to \infty} \left( \frac{1}{t} \ln \frac{p(\overline{\Lambda}_t > 0)}{p(\overline{\Lambda}_t < 0)} \right) = \frac{1}{t} \ln \left\langle \exp(\overline{\Lambda}_t) \right\rangle_{\overline{\Lambda}_t < 0} \qquad (12)$$

and,

$$\lim_{t \to \infty} \left( \frac{1}{t} \ln \frac{p(\bar{\Omega}_t < 0)}{p(\bar{\Omega}_t > 0)} \right) = \frac{1}{t} \ln \left\langle \exp(-\bar{\Omega}_t t) \right\rangle_{\bar{\Omega}_t > 0} \qquad (13)$$

where the notation $\langle ... \rangle_{\bar{\Lambda}_t < 0}$ and $\langle ... \rangle_{\bar{\Omega}_t > 0}$ are used to denote conditional ensemble averages. We note that in obtaining (12), it is assumed that equation (2) is valid for all observable values of A. In case this is not true in all systems [40], we also test (2) directly [41]. In figure 1a) a direct test of equations (12) & (13) are plotted using data from the steady state simulations with $\varepsilon = 0.1$. We observe that equation (13) (ES) converges on the simulation time scale while equation (12) (GC) does not. This is shown in more detail at the longest averaging time in figure 1b) where (2) is tested for $\langle \bar{\Lambda}_t \rangle / 2 \leq A \leq -\langle \bar{\Lambda}_t \rangle / 2$ (or $-\frac{1}{2} \leq p \leq \frac{1}{2}$, where $p = -A / \langle \bar{\Lambda}_t \rangle$ see [40]). At this time it can be seen that equation (3) (ES) has largely converged while equation (2) (GC) has not. A more rapid decay for the large field case of $\varepsilon = 0.43$ may be seen in figure 2a). For this strongly driven system there is one pair of contracting/expanding exponents, and the other pair which are both contracting (see Table 1). The postulate of reference [11] requires that $X = 0.5$ under these conditions and the resulting prediction of equation (5) (GCX) and (6) (GX) may be seen in figure 2a). A more detailed comparison to equations (2) (GC), (3) (ES), (5) (GCX) & (6) (GX) at the longest averaging time (t=1000) are shown in figure 2b). Here $0.156 \langle \bar{\Lambda}_t \rangle \leq A \leq -0.156 \langle \bar{\Lambda}_t \rangle$ (or $-0.156 \leq p \leq 0.156$, see [40]). Data at t=600 is also shown to demonstrate that convergence to the long time behaviour has occurred, within the limits of error of the data. Clearly good numerical agreement with the predictions of (3) (ES) are observed and there is no indication that a factor of $X = 0.5$ should be included in equation (5) or (6). At this high field strength equation (2) (GC) converges on a reasonable time scale. At field strengths considerably lower than $\varepsilon = 0.1$ it may be that equation (2) (GC) never converges. To illustrate this we show a comparison of equations (12) (GC) & (13) (ES) for the very low field strength of $\varepsilon = 0.01$ in figure 3. Again convergence is observed for equation (13) but in the case of equation (12) there is no evidence of convergence on the longest simulation time scale that we have computed. This longest time is approximately 100 times the characteristic microscopic relaxation time (ie Maxwell time) for the system.

Some equilibrium and steady state probability distributions for a related, but different system are shown in reference [34]. This suggests that the distributions for weakly driven systems are similar to the equilibrium distributions and span the full phase space, but this is not the case for strongly driven systems. We present similar data for our system. Figure 4a) and 4b) represent the equilibrium distributions projected onto the $qp$ and $q\alpha_5$ planes, respectively. These projections for steady state systems with $\varepsilon=0.1$ and $\varepsilon=0.43$ are shown in Figures 4c)-f). At $\varepsilon=0.43$ the phase space is no longer filled, and the system should not be considered to be transitive, yet as shown above the FR does not change. In Figure 5, initial points of trajectory segments with positive and negative values of $\bar{\Omega}_t$, averaged over t = 1000 are shown. This demonstrates that the attractor is chaotic with nearby points generating very different values of $\bar{\Omega}_t$. Despite the dominantly negative Lyapunov spectrum the distribution of positive and negative dissipation points is very similar. We note that equal numbers of trajectories with positive and negative values of $\bar{\Omega}_t$ were selected for this figure even though the proportion of initial points that produce negative $\bar{\Omega}_t$ will be very small at this field and value of t.

From these results, it can be concluded that there is no sudden change in the applicability of the fluctuation relations when traversing from a steady state spectrum with 2 positive and 2 negative exponents to a spectrum with 1 positive and 3 negative exponents.

The derivation of (3) (ES) assumes that the statistics of the time-averaged properties sampled from an initial distribution that has then evolved towards an attractor will match those of segments sampled from the steady state attractor in the infinite time limit. This will be true if only a single steady state can be identified, so we have restricted ourselves to this case here.

For temperature controlled dynamics equations (2) and (3) are different, and equation (2) may be used to derive results in contradiction to well established Green-Kubo formula [33]. The derivation of equation (2) leads to the expectation that it should break down when the transitive property is lost but we fail to observe this. There is no known system where equation (2) can be numerically observed to converge faster than equation (3) for steady state temperature controlled dynamics. This suggests that the observed convergence of equation (2) at high field strengths is a result of the fluctuations in the phase space compression factor (8) being strongly correlated to the fluctuations in the dissipation function (11) [33] and is not a result of temperature controlled dynamics being Anosov-like.

**Conclusions**

We have performed a number of new tests of both the Evans-Searles and the Gallavotti-Cohen fluctuation relations (ESFR and GCFR respectively). The system involves a triply thermostatted model of heat flow. The model is unusual in that the Lyapunov exponents are "soft": we can observe a change in sign of one of the four nontrivial Lyapunov exponents at fields strengths (ie temperature gradients) that are still small enough to observe fluctuations which, were they to continue for a very long time, would be in violation of the Second Law of Thermodynamics. It is these "Second Law violating" fluctuations that are the subject of the various fluctuation Relations.

Near equilibrium the ESFR (2) is verified by the simulation data while the GCFR (3) is not confirmed by the data. The data shown for the integrated version of the steady state Fluctuation Relations is really very stark. For the weak fields studied in that figure the steady state ESFR (2) converges but even at 100 times the Maxwell relation time the GCFR (3) has still failed to converge. In fact the data in the graph could imply that the GCFR never converges regardless of how large the averaging time.

Far from equilibrium where a positive exponent in one of these conjugate pairs becomes negative, we test a conjecture by Gallavotti and coworkers [11, 32]. They conjectured that where the number of nontrivial Lyapunov exponents that are positive becomes less than the number of such negative exponents, then the form of the GCFR needs to be corrected. We show that there is no evidence for this conjecture in our numerical data. In fact as the field increases, the uncorrected form of the GCFR appears to become more accurate. The reason for this observation is likely to be that as the field increases, the argument of the GCFR more and more accurately approximates the argument of the ESFR. Since the ESFR works for arbitrary field strengths, the uncorrected GCFR appears to become ever more accurate as the field increases. The final point of evidence against the conjecture is that when the smallest positive exponent changes sign the conjecture predicts a discontinuous change in the "correction factor" for GCFR. We see no evidence for a discontinuity at this field strength in either the GCFR or in the ESFR. We only see a gradual improvement of degree of agreement as the field increases.

We note that similar behavior has recently been observed by Tempatarachoke [39] who have also verified (3) for a system that is far from equilibrium and shown (numerically) that transitivity is unnecessary.

**Acknowledgments**

The authors wish to thank the Australian Research Council, The Queensland Parallel Computing Facility and the Australian National Facility for support of this work. The Erwin Schrödinger Insitute is thanked for their support of the "Workshop on Stochastic and Deterministic Dynamics in Equilibrium and Nonequilibrium Systems" where the model studied in this work was introduced to us. We also thank Lamberto Rondoni for his useful comments.

**Figure Captions**

Figure 1. a) Semi logarithmic plot displaying a direct test of Eq. 12 & 13 with $\varepsilon = 0.1$. The diamonds are the estimates of the ratio $p(\bar{\Omega}_t < 0)/p(\bar{\Omega}_t > 0)$ as a function of time and the solid line is the estimate of the average $\langle \exp(-\bar{\Omega}_t t) \rangle_{\bar{\Omega}_t > 0}$. The stars are the estimates of the ratio $p(\bar{\Lambda}_t > 0)/p(\bar{\Lambda}_t < 0)$ and the dashed line is the estimate of the average $\langle \exp(\bar{\Lambda}_t t) \rangle_{\bar{\Lambda}_t < 0}$.

b) A direct test of equations (2) & (3) for the integration time $t = 10\,000$. The solid line is the theoretical prediction.

Figure 2a) Same as figure 1a) but for $\varepsilon = 0.43$. The additional dotted line is the estimate of $\langle \exp(-X\bar{\Omega}_t t) \rangle_{\bar{\Omega}_t > 0}$ with X = 0.5 which tests equation (6) and the additional dash dotted line is the estimate of $\langle \exp(X\bar{\Lambda}_t t) \rangle_{\bar{\Lambda}_t < 0}$ with X = 0.5 which tests Eq 5. For clarity the stars and the dashed and dash-dotted line data have been divided by a factor of 10 to move them one decade down the ordinate axis.

b) A direct test of equations (2), (3), (5) and (6) for the integration time $t = 1000$. The solid line is the theoretical prediction of (2) or (3), while the dashed line is the prediction of (5) or (6) with X = 0.5. Data for t = 600 is also shown to demonstrate the convergence.

Figure 3. Linear plot displaying a direct test of equations (12) & (13) for the very weak field of $\varepsilon = 0.01$. Other details are as in figure 1a).

Figure 4. Projections of equilibrium and steady state distributions onto the *qp* plane and the $q\alpha_5$ planes for system 2. $\varepsilon = 0$ for a) & b), $\varepsilon = 0.1$ for c) & d) and $\varepsilon = 0.43$ for e) & f). Each projection has $5 \times 10^4$ points plotted.

Figure 5. Starting points for trajectory segments with positive and negative dissipation over a duration of t=1000 for system 2 with $\varepsilon = 0.43$. There are 637 negative and 637 positive starting points shown, even though the probability of observing a positive trajectory segment is far higher than that of observing a negative segment.

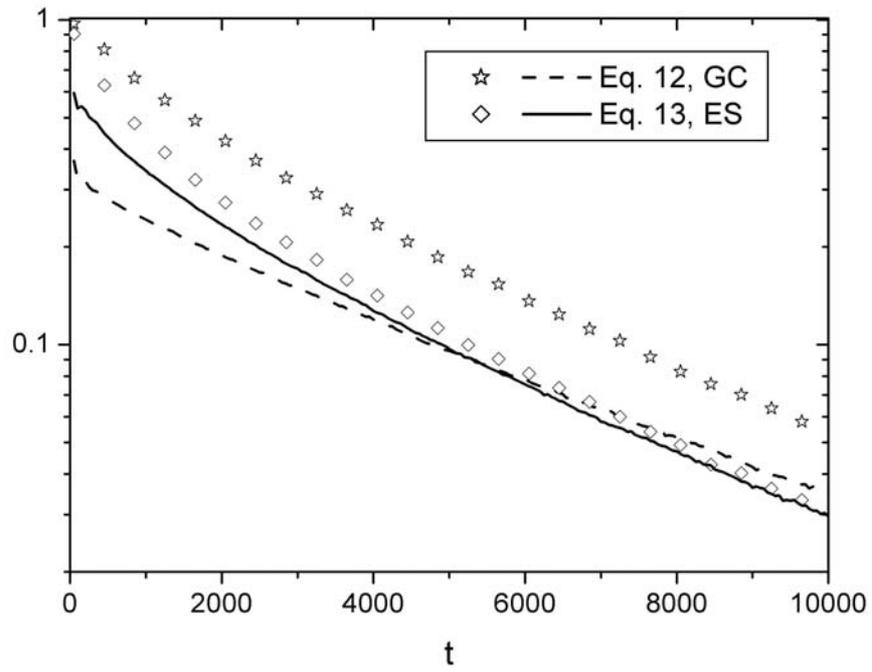

Figure 1a)

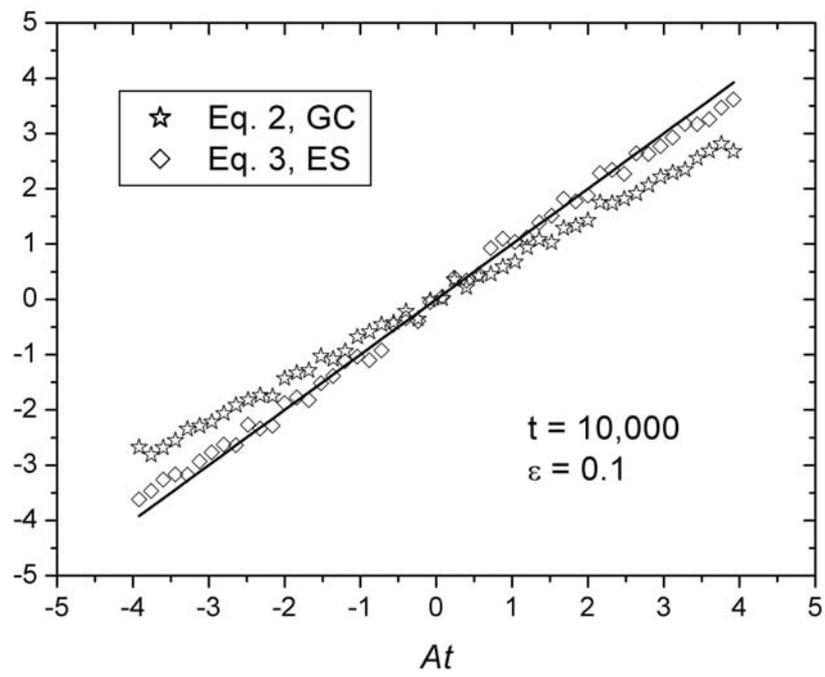

Figure 1b)

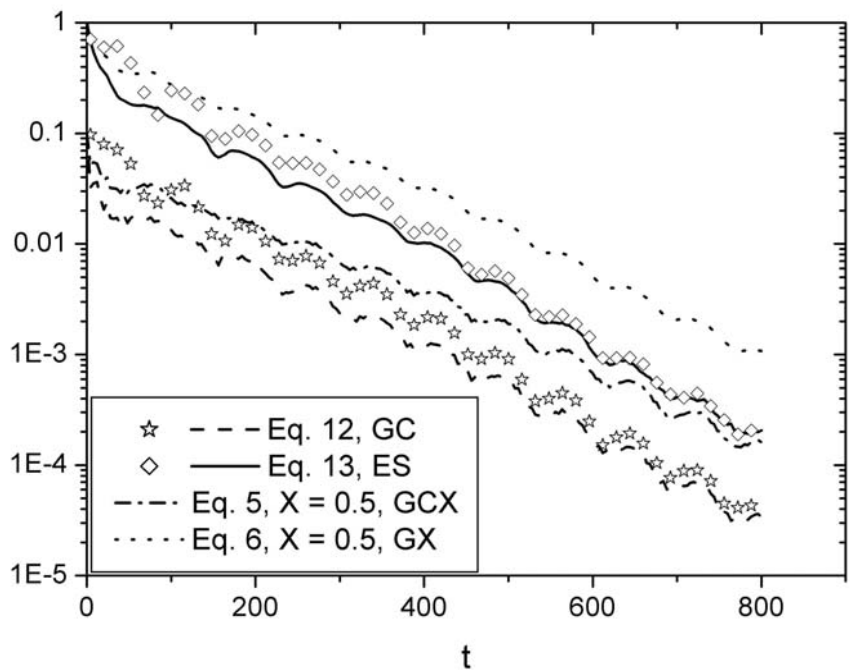

Figure 2a)

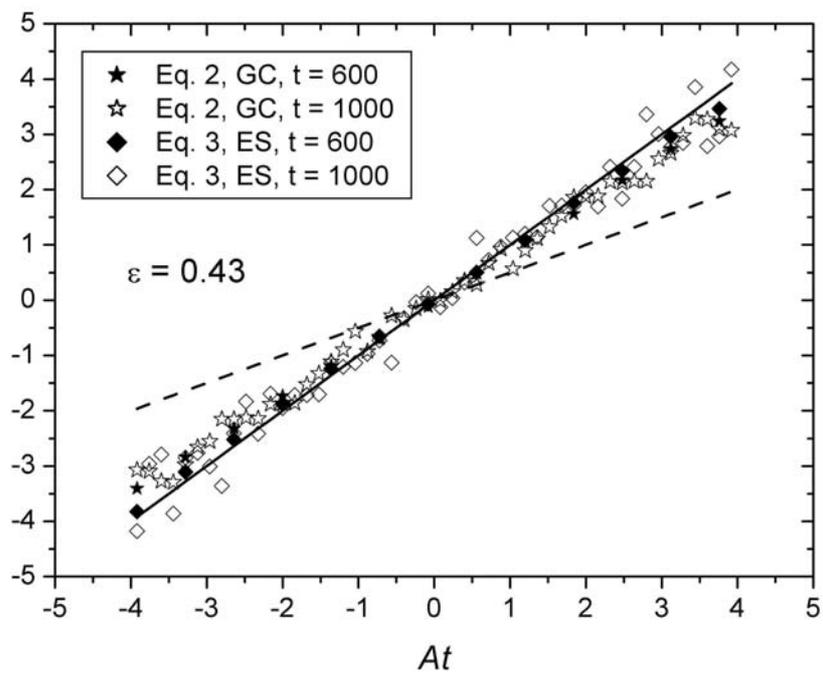

Figure 2b)

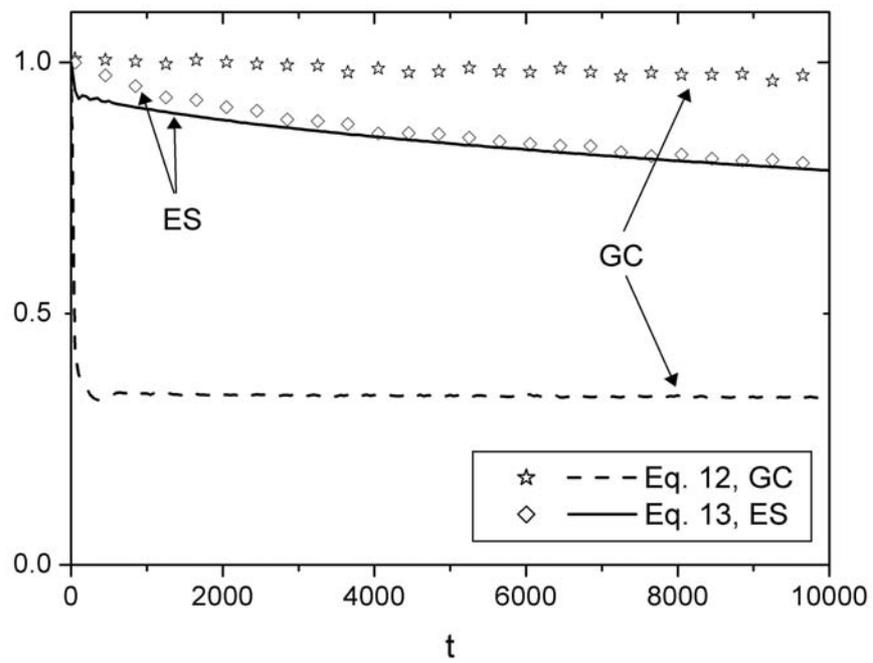

Figure 3)

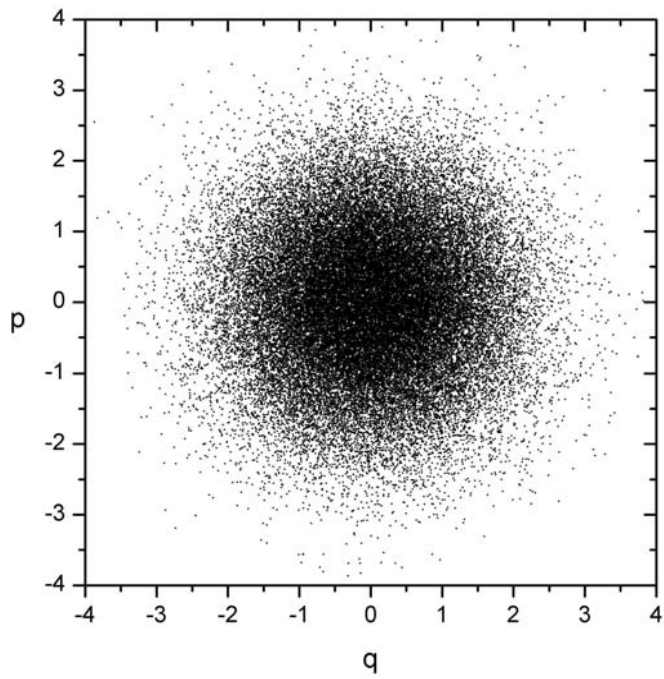

Figure 4a)

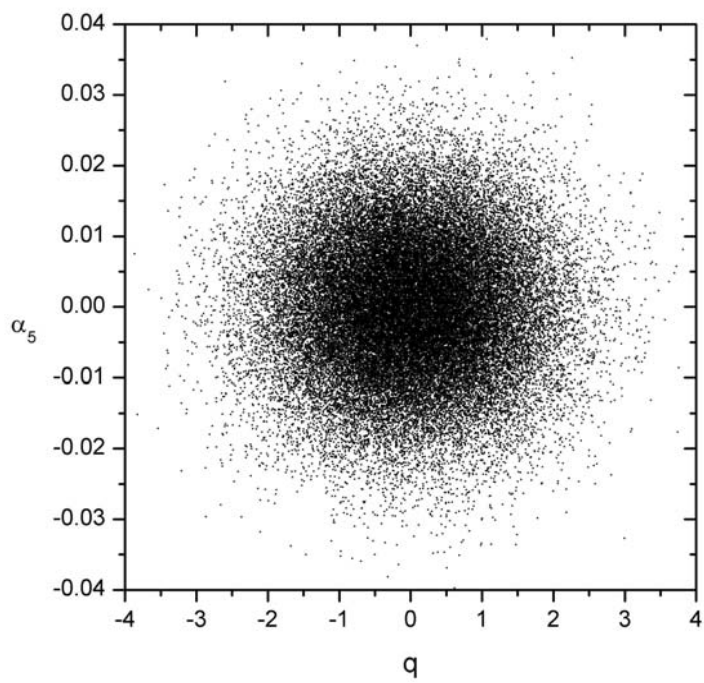

Figure 4b)

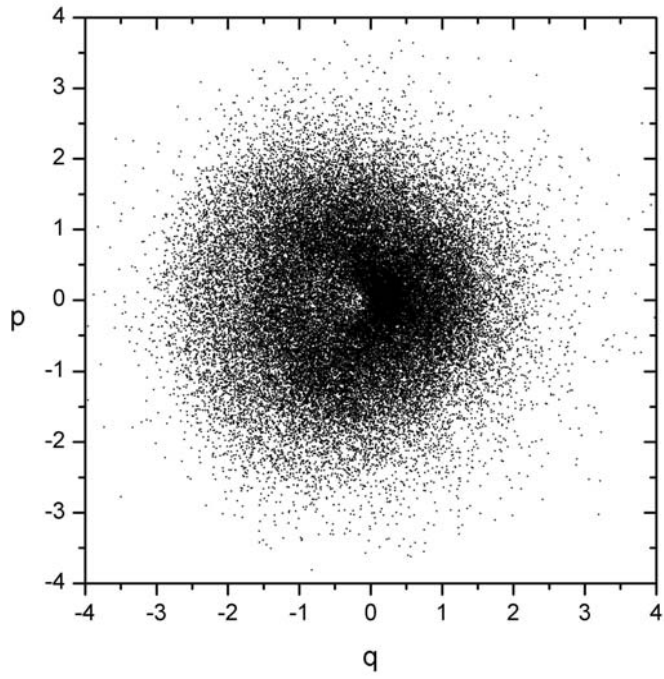

Figure 4c)

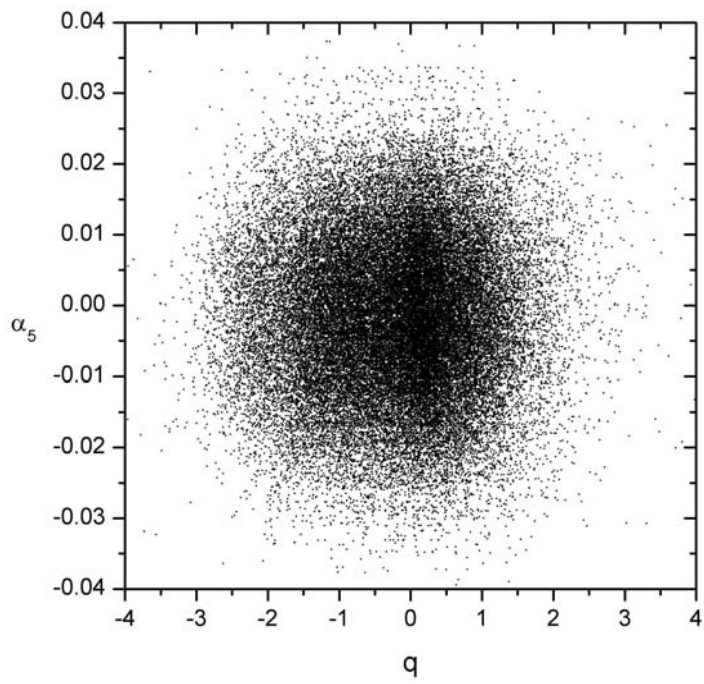

Figure 4d)

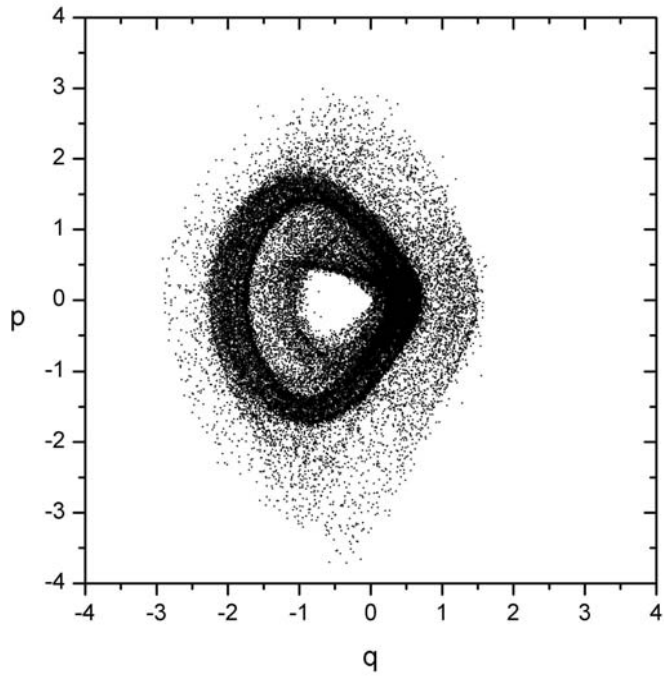

Figure 4e)

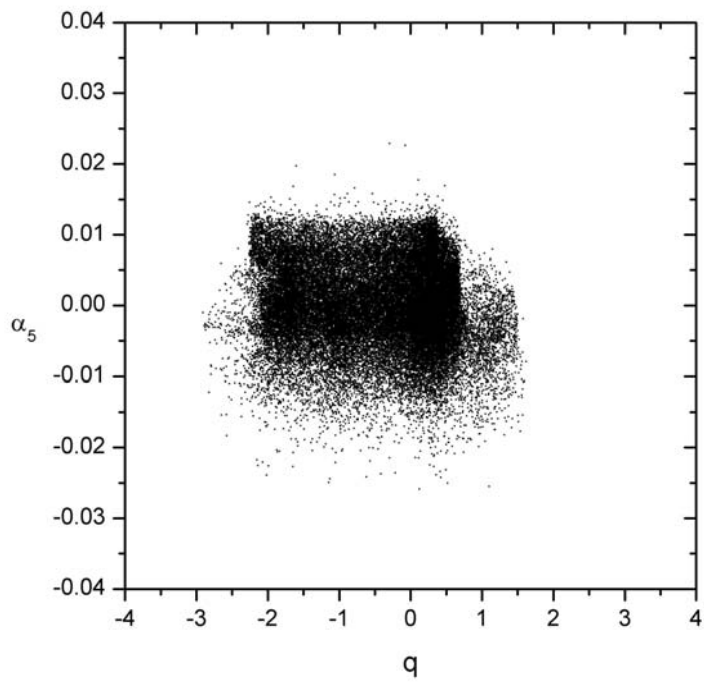

Figure 4f)

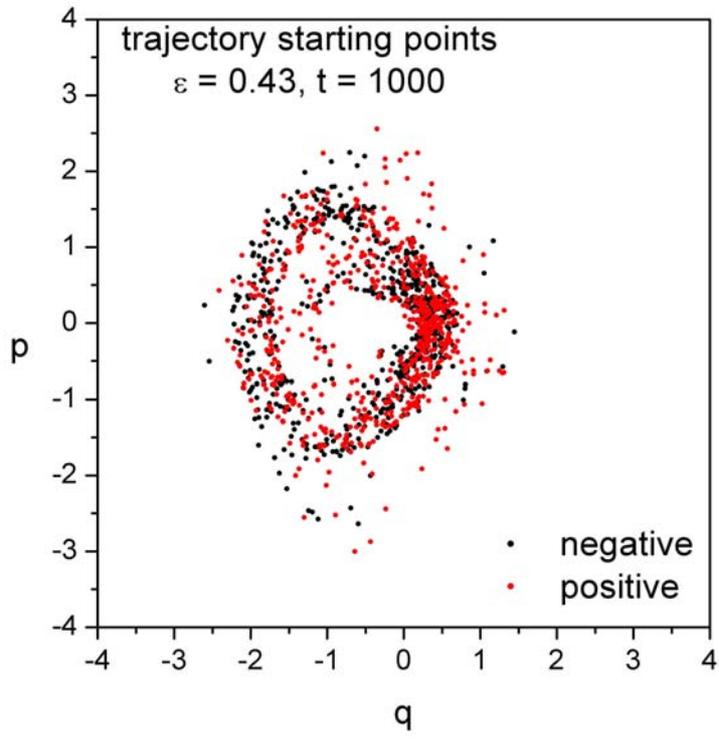

Figure 5.

# References

[1] D. J. Searles and D. J. Evans, Phys. Rev. E **60**, 159 (1999)

[2] G. Ayton, D. J. Evans, and D. J. Searles, J. Chem. Phys. **115**, 2033 (2001)

[3] D. J. Evans and D. J. Searles, Phys. Rev. E **52**, 5839 (1995)

[4] D. J. Evans and D. J. Searles, Phys. Rev. E **50**, 1645 (1994)

[5] D. J. Evans and D. J. Searles, Phys. Rev. E **53**, 5808 (1996)

[6] D. J. Searles and D. J. Evans, J. Chem. Phys. **113**, 3503 (2000)

[7] D. J. Searles and D. J. Evans, J. Chem. Phys. **112** (2000)

[8] D. J. Searles and D. J. Evans, Int. J. Thermophysics **22**, 123 (2001)

[9] D. J. Evans, D. J. Searles, and E. Mittag, Phys. Rev. E **63**, 051105/1 (2001)

[10] D. J. Searles, G. Ayton, and D. J. Evans, AIP Conference Series **519**, 271 (2000)

[11] F. Bonetto, G. Gallavotti, and P. L. Garrido, Physica D **105**, 226 (1997)

[12] F. Bonetto, N. I. Chernov, and J. L. Lebowitz, Chaos **8**, 823 (1998)

[13] L. Biferale, D. Pierotti, and A. Vulpiani, J. Phys. A **31**, 21 (1998)

[14] F. Bonetto and J. L. Lebowitz, Phys. Rev. E **64**, 056129 (2001)

[15] S. Sasa, (arXiv:nlin.CD/0010026, 2000).

[16] S. Lepri, R. Livi, and A. Politi, Phys. Rev. Lett. **78**, 1896 (1997)

[17] A. Baranyai, J. Chem. Phys. **119**, 2144 (2003)

[18] M. Schmick and M. Markus, Phys. Rev. E **70**, 065101/1 (2004)

[19] M. Dolowschiak and Z. Kovacs, Phys. Rev. E **71**, 025202/1 (2005)

[20] F. Zamponi, G. Ruocco, and L. Angelani, J. Stat. Phys. **115**, 1655 (2004)

[21] D. M. Carberry, J. C. Reid, G. M. Wang, E. M. Sevick, D. J. Searles, and D. J. Evans, Phys. Rev. Lett. **92**, 140601 /1 (2004)

[22] J. C. Reid, D. M. Carberry, G. M. Wang, E. M. Sevick, D. J. Evans, and D. J. Searles, Phys. Rev. E **70**, 016111/1 (2004)

[23] G. M. Wang, E. M. Sevick, E. Mittag, D. J. Searles, and D. J. Evans, Phys. Rev. Lett. **89**, 050601/1 (2002)

[24] N. Garnier and S. Ciliberto, Phys. Rev. E **71**, 060101/1 (2005)

[25] D. J. Evans and D. J. Searles, Ad. Phys. **51**, 1529 (2002)

[26] D. J. Evans, E. G. D. Cohen, and G. P. Morriss, Phys. Rev. Lett. **71**, 2401 (1993)

[27] G. Gallavotti, Mathematical Physics Electronic Journal **1**, 12pp (1995)

[28] G. Gallavotti and E. G. D. Cohen, J. Stat. Phys. **80**, 931 (1995)

[29] G. Gallavotti and E. G. D. Cohen, Phys. Rev. Lett. **94**, 2694 (1995)

[30] D. J. Evans and G. P. Morriss, *Statistical Mechanics of Non-equilibrium Liquids* (Academic, London, 1990).

[31] D. J. Searles, L. Rondoni, and D. J. Evans, in preparation (2005)

[32] A. Giuliani, F. Zamponi, and G. Gallavotti, J. Stat. Phys. **119**, 909 (2005)

[33] D. J. Evans, D. J. Searles, and L. Rondoni, Phys. Rev. E **71**, 056120 (2005)

[34] W. G. Hoover and C. G. Hoover, Condensed Matter Physics **8**, 247 (2005)

[35] W. G. Hoover, C. G. Hoover, H. A. Posch, and J. A. Codelli, Communications in Nonlinear Science and Numerical Simulation, in press (2005)

[36] D. J. Searles, D. J. Evans, and D. J. Isbister, Chaos **8**, 337 (1998)

[37] S. Sarman, D. J. Evans, and G. P. Morriss, Phys. Rev. A **45**, 2233 (1992)

[38] C. Dellago, H. A. Posch, and W. G. Hoover, Phys. Rev. E **53**, 1485 (1996)

[39] P. Tempatarachoke, in *School of Physical, Environmental and Mathematical Sciences,* (The University of New South Wales at The Australian Defence Force Academy, Canberra, 2004), p. 203.

[40] Determination of the value of $A^*$ is usually very difficult, however statements of the GCFR often specify that $1 \leq p^* \leq \infty$, where $p^* = -A^*/\langle \overline{\Lambda}_t \rangle$, and $\langle \overline{\Lambda}_t \rangle$ is

the mean value of $\bar{\Lambda}_t$ (see [32]), which has a negative value. Furthermore, a new treatment of systems with Nosé-Hoover thermostats suggests that equation (2) might only be valid in the range $|p| \leq 1$ for these systems ( F.Bonetto, G.Gallavotti, A.Giuliani, F.Zamponi, cond-mat/0507672). Therefore, if it applies, the GCFR should be valid for *at least* $|p| \leq 1$, or equivalently $|A| \leq |\langle \bar{\Lambda}_t \rangle|$.

[41] It is also possible to obtain modified versions of (12) where only $A \in (-A^*, A^*)$ is considered (i.e. $\lim_{t \to \infty} \left( \frac{1}{t} \ln \frac{p(0 < \bar{\Lambda}_t < A^*)}{p(-A^* < \bar{\Lambda}_t < 0)} \right) = \frac{1}{t} \ln \langle \exp(\bar{\Lambda}_t t) \rangle_{-A^* < \bar{\Lambda}_t < 0}$). However, we might expect the numerical results for many systems would be very similar to (12) in the long time limit since the contribution to $\langle \exp(\bar{\Lambda}_t t) \rangle_{\bar{\Lambda}_t < 0}$ of values of $\bar{\Lambda}_t < -A^*$ will be small at large t.